# Unconventional magnetic anisotropy in one-dimensional Rashba system realized by adsorbing Gd atom on zigzag graphene nanoribbons

Zhenzhen Qin[1], Guangzhao Qin[2], Bin Shao[3], Xu Zuo[1*]

[1]*College of Electronic Information and Optical Engineering,*

*Nankai University, Tianjin, 300350, China*

[2] *Institute of Mineral Engineering, Division of Materials Science and Engineering,*

*Faculty of Georesources and Materials Engineering,*

*RWTH Aachen University, Aachen 52064, Germany*

[3] *Bremen Center for Computational Material Science,*

*Bremen University, Am Fallturm 1, Bremen 28359, Germany*

___________________

* Author to whom all correspondence should be addressed: xzuo@nankai.edu.cn.

# ABSTRACT

The Rashba effect, a spin splitting in electronic band structure, attracts much attention for the potential applications in spintronics with no requirement of external magnetic field. Realizing one-dimensional (1D) Rashba system is a big challenge due to the difficulties of growing high-quality heavy-metal nanowires or introducing strong spin-orbit coupling (SOC) and broken inversion symmetry in flexible materials. Here, based on first-principles calculations, we propose a pathway to realize the Rashba spin-split by adsorbing Gd atom on zigzag graphene nanoribbons (Gd-ZGNR) and further investigate the magnetic anisotropy energy (MAE). Perpendicular MAE and unconventional MAE contributions in *k*-space are found in the self-assembled Gd-ZGNR system, which presents a remarkable Rashba effect (the estimated strength is 1.89 eV Å) due to the strong SOC (~65.6 meV) and the asymmetric adsorption site at nanoribbons edge. Moreover, first-order MAE is connected to the intrinsic Rashba effect beyond the traditional second-order MAE, which is confirmed based on the analysis of electronic structures perturbed with SOC in comparison with metastable Gd-ZGNR at central symmetric adsorption site. The dependence on the ribbon width of first-order MAE as well as Rashba effect in Gd-ZGNRs are also examined. This work not only opens a new gate for designing 1D Rashba system but also provides insight into the unconventional MAE due to the intrinsic Rashba effect, which would be of great significance for searching Majorana fermions and promoting the potential applications in spintronics.

## I. INTRODUCTION

The Rashba effect, a spin splitting in electronic band structure, attracts much attention for the potential applications in spintronics with no requirement of magnetic field [1,2]. It was discovered firstly in bulk non-centrosymmetric wurtzite semiconductors [3,4] and further extended to two-dimensional (2D) systems [5-8] due to the spin-orbit coupling (SOC) and the broken inversion symmetry. Although Rashba effects mostly exist in 2D surfaces or interfaces, intensive attempts have been made to one-dimensional (1D) systems with large spin splitting for the distinctive advantages, such as the merits in the downsizing of devices, manipulating spin current [9], study for helical liquid states [10, 11] and the search for Majorana fermions [12,13]. Whereas, to realize 1D Rashba systems with strong SOC has been a long-time challenge because of the small intrinsic spin-split in common 1D carbon nanotubes or semiconductor quantum wires. Recently, some experimental groups attempt to achieve the Rashba effect in pure heavy-metal (such as Au [14] and Pt [15]) nanowires artificially grown on a silicon surface. Even so, the current 1D Rashba systems still face the limitation of growth quality and structural flexibility in practice.

Note that 1D graphene nanoribbons (GNR) with defined width can be realized either by cutting mechanically exfoliated graphene and patterning by lithographic techniques [16, 17] or by directly grown on hexagonal boron nitride using chemical vapor deposition [18], which overcome the difficulties of growth technology and have great advantages on the mechanical and thermal properties. The retained problem is how to introduce strong SOC and broken inversion symmetry in such GNR systems, which is the key requirement of Rashba effect. In general, foreigner atoms combine with graphene system via adsorption has been proved to be an effective way to control the spins of electrons or SOC strength of graphene-based devices [19-22]. Thus, the GNR could be a good candidate matrix material by introducing the essential factors of Rashba effect. On the other hand, among the 1D form of graphene [23], zigzag graphene nanoribbon (ZGNR) exhibit an unique edge states with opposite spin polarization [24-26], and the electronic and magnetic properties of nano-materials are typically much enhanced compared to normal materials due to the quasi-one-dimensional features and edge states [25,27]. Considering the unique edge states and magnetic planes of ZGNR, which serve as a source of exchange field, the magnetic properties of impurity on ZGNR may present diverse phenomenon compared to the freestanding graphene. However, the research on ZGNR-based materials has just started [28-33] either in experimental or theoretical levels, while whether the Rashba effect can be realized in such systems or not is still not investigated.

In this paper, based on first-principles calculations we successfully predict a self-assembled Rashba system by adsorbing Gd atom on zigzag graphene nanoribbons (Gd-ZGNR). Beyond that, the electronic and magnetic properties (especially magnetic anisotropy) of Gd-ZGNR are fully investigated. The orbital hybridization, magnetic moment and magnetic anisotropy in $k$-space are found to depend on the Gd adsorption sites. In particular, a remarkable Rashba effect is demonstrated in the self-assemble Gd-ZGNR system, which presents an unconventional MAE contributions in $k$-space. Beyond the traditional second-order MAE, we establish a microscopic connection between Rashba spin splitting and first-order MAE by analyzing the electronic structures perturbed with SOC, comparing with metastable Gd-ZGNR. In addition, the real second-order MAE of Gd-ZGNR systems is explained based on the second-order perturbation theory. Finally, the nature of Rashba spin-split in Gd-ZGNR are confirmed by examining more systems with different widths. The Rashba effect predicted in Gd-ZGNR has obvious advantages over the already known 1D Rashba systems [14,15] in terms of the combination of heavy element with versatile graphene nanoribbon substrate in a self-assembled pattern. Our study not only provides deep understanding of the Rashba nature in Gd-ZGNR but also contributes to the development of ZGNR-based spintronics and high density data storage.

## II. COMPUTATIONAL DETAILS

Spin-polarized density functional calculations are performed using the projected augmented wave (PAW) [34] method as implemented in the Vienna *ab-initio* simulation package (VASP) [35]. Exchange-correlation energy functional is treated in the Perdew-Burke-Ernzerhof of generalized gradient approximation (GGA-PBE) [36]. The valence electron configurations of Gd are considered as $4f^7 5d^1 6s^2$. The wave functions are expanded in plane wave basis with a kinetic energy cutoff of 450 eV. The partially filled and strongly correlated localized *f* orbitals of Gd atoms are treated using the GGA+U formalism [37,38] with $U_f$= 8 eV [39]. The systems of *n*-ZGNRs (the width *n*=4, 5, 6, 7, 8, 9, 10, 12) with Gd atom adsorbed are studied here. To avoid interaction between infinite nanoribbons made by periodic boundary conditions, the vacuum region is set as 20 Å in the *y* direction and 15 Å in the *z* direction (the nanoribbon along the *x* direction is treated as periodic and infinite). To saturate the edge C dangling bonds, the nanoribbons are passivated by hydrogen atoms. Brillouin zones (BZ) are sampled using 15×1×1 Monkhorst-Pack grids for the geometry optimization. The Gd adatoms are allowed to relax along the *z* direction and the C atoms in are allowed to relax along all directions, until the force on each atom is less than 0.02 eV/Å. The convergence criterion of the self-consistent-field (SCF) calculations is $10^{-6}$ eV. In the self-consistent potential and total energy calculations, a set of 25×1×1 *k*-point sampling is used for BZ integration. For band structure calculations 60 uniform *k* points along the one-dimensional BZ is used. The magnetic anisotropy energy (MAE) is calculated following the Force theorem [40]. The dipole correction [41] is included in the calculations.

## III. RESULTS AND DISCUSSION

ZGNR is a semiconductor with two electronic edge states, which are ferromagnetically ordered in one edge, but antiferromagnetically coupled to each other [42,43]. Our calculations show that the antiferromagnetic (AFM) state has energy by Δ*E*=53.86 meV per edge carbon atom lower compared to the unpolarized case and 14.26 meV lower than the ferromagnetic (FM) state for 4-ZGNR (width $n$=4). The obtained values are in good agreement with previous results [32]. The calculated electronic properties of pure 4-ZGNR are in good agreement with previous reports [28-32] as shown in Fig. S1 in the Supplemental Material. Specifically, the spin moments are mainly distributed at the edge carbon atoms and slower decay towards the center of the ribbon. The magnetic moment fluctuation across the ribbon arises from quantum interference effects caused by nanoribbon edges. Due to the topology of the lattice, the atoms of the two edges belong to different sublattices of the bipartite graphene lattice. The spin density on the C atoms on one edge are anti-aligned to the spin density on the opposite edge, and the polarizations of neighboring sites belonging to different sublattices are also opposite. The band structures of ZGNR decorated with up/down C-$p_z$ orbitals are presented in Fig. S1(c) of the Supplemental Material.

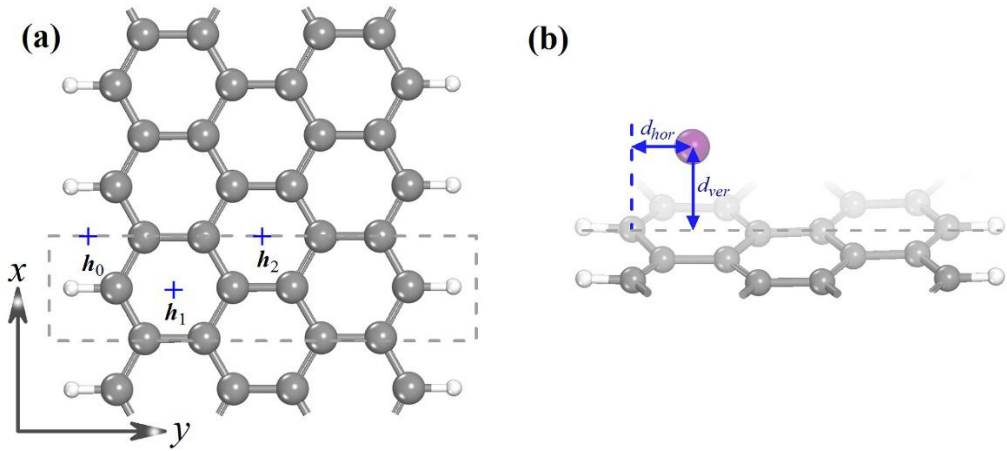

FIG. 1. (Color online) (a) Structure of H-passivated 4-ZGNR, where the gray and white balls represent C and H atoms, respectively. The possible hollow sites for Gd adsorption on 4-ZGNR ($h_1$, $h_2$ and the perturbed hollow-like site of the edge $h_0$) are labeled on site. The unit cell of 4-ZGNR is indicated by dashed lines. (b) Side view of the favorable $h_1$-Gd-ZGNR configuration, where the purple ball represents Gd adatom. The $d_{ver}$ is defined as the average vertical distance between adatom and ZGNR plane, and $d_{hor}$ is the horizontal distance between adatom and the left edge Carbon atom.

## A. Stability of Gd adsorbed ZGNR

Here, we first consider the 4-ZGNR system, which was mainly adopted in previous studies to describe the electronic properties of a Co adatom on nanorribon [32]. As shown in Fig. 1(a), the case of Gd-4ZGNR corresponds to a coverage of one adatom per 32 C atoms. Then, we examine the most favorable structure for Gd adsorbed 4-ZGNR. Noted that the hollow site is more preferred than the top or bridge site for Gd adsorbed on the hexagonal lattice of graphene [44]. Unlike graphene, due to the broken symmetry along the transverse direction, three inequivalence possible hollow sites are considered ($h_0$; $h_1$; $h_2$, as marked in Fig. 1(a)). As for the binding energy of a Gd adatom on ZGNR, one can use the formula

$$E_b = E_{Gd} + E_{ZGNR} - E_{Gd+ZGNR} \tag{1}$$

where $E_{Gd}$, $E_{ZGNR}$, and $E_{Gd-ZGNR}$ is the total energy of an isolated single Gd adatom, ZGNR and Gd-ZGNR system, respectively. From Table 1, we conclude that the edge hollow $h_1$ is most preferred adsorption site for Gd-4ZGNR. Similarly, other elements(V [31] or Co [32]) also actually tend to adsorb on the edge hollow site of ZGNR. Except for $h_1$ and $h_2$, Gd adatom in $h_0$ site inevitably lead to the unreasonable structural distortion of ZGNR plane after full relaxation, which we will discuss shortly in following text.

TABLE I. Binding energies and the total magnetic moment excluding Gd-4f to various adsorption positions of Gd-4ZGNR. The structural parameters are also provided, including the $d_{ver}$ after relaxation and $d_{hor}$ after (and before) relaxation.

|  | $h_2$ | $h_1$ | $h_0$ |
|---|---|---|---|
| $E_b$ (eV) | 1.57 | 2.04 | 1.61 |
| $\mu_{\bar{f}}(\mu_B)$ | 1.07 | 2.20 | 2.57 |
| $d_{ver}$ (Å) | 2.14 | 2.12 | 2.03 |
| $d_{hor}$ (Å) | 3.561 (3.559) | 1.393 (1.419) |  |

The geometric structures of $h_1$/$h_2$-Gd-ZGNR are further discussed. As shown in Fig. 1(b), the $d_{ver}$ and $d_{hor}$ are defined as the vertical distance between adatom and ZGNR plane, and the horizontal distance between adatom and the left edge Carbon atom, respectively. For Gd adsorbed $h_1$ and $h_2$ site, the $d_{ver}$ are nearly equal (2.12 and 2.14 Å) after full relaxation. While the $d_{hor}$ changes from 1.419 Å to 1.393 Å (the variation ~ 0.026 Å) at $h_1$ site; the $d_{hor}$ changes little (~ 0.002 Å) at $h_2$ site. The variation of $d_{hor}$ after/before relaxation reveals that large structural distortion can be induced by the asymmetry adsorption. Following the method used in Ref. 45, the diffusion barrier ($E_{bar}$) can be obtained from the energy difference between the hollow and bridge sites according to the symmetry. The $E_{bar}$ is estimated to be 0.21 eV, which is consistent with previous theoretical result (0.23 eV) [46]. The diffusion barrier of Gd on ZGNR is two times larger than that of most group I-III elements (~ 0.1 eV) on graphene [45]. In addition, previous experimental study reported that Gd cannot easily form clusters unless at high temperatures due to the large diffusion barriers [46]. Hence, we can conclude that it is difficult for Gd atoms to form clusters on ZGNR due to the large diffusion barrier. Although current experimental researches only focus on Gd adsorbed graphene [46], it can be expected that the system of Gd adsorbed on zigzag graphene nanoribbons (ZGNR) with deifined widths can be readily prepared by directly grown on hexagonal boron nitride using chemical vapor deposition [18]. Note that for neutral Gd atom, there are seven $f$ electrons in the ground state. The half-filled $f$-electron configuration is energetically quite stable, revealing that $f$-electron magnetic moment (the calculated value is 6.97 $\mu_B$) is not a determining factor in intra-atomic exchange of systems. Thus, in Table I, we only present the total magnetic moments with excluding the $f$-orbital, and so does the subsequent analysis.

B. Electronic and magnetic properties of Gd-ZGNR

We provide the orbital-projected density of states (DOS) of Gd-ZGNR at $h_1$ site, and give that of $h_2$-Gd-ZGNR for comparison, as shown in Fig. S2 in the Supplemental Material. It is found that the orbital splitting of Gd-5$d$ levels are almost no difference in the two adsorbed cases: The down-spin density of Gd-$d$ orbitals are all empty, while the up-spin density reveals that $d_{xy}$, $d_{yz}$, $d_{xz}$ orbitals are nearly unoccupied, which lie higher in energy than the obvious partially filled $d_{x^2-y^2}$ and $d_{z^2}$ orbitals. The strong hybridization between the Gd $sd$-orbital and C-$p_z$ orbital is confirmed by the band dispersions and global DOS (Fig. 2(a) and 2(b)). It can be clearly seen that the electronic property of Gd-ZGNR in $h_1$ and $h_2$ is completely different (metallic of $h_1$-Gd-ZGNR vs. half-metallic of $h_2$-Gd-ZGNR), which is mainly dominated by the different hybridization between Gd-5$d$ and C-$p_z$ orbitals. Similarly, the spin density of Gd-ZGNR can be largely affected by the adsorption location of Gd (Fig. 2(c) and 2(d)). Actually, the localized edge states is determined by the adsorption location of Gd, which would play a key role on the electronic structure of Gd-ZGNR at different adsorption sites. It is known that the C atoms at each edge site of ZGNR belonging to different sublattices, where the spin polarizations are remaining opposite due to the topology of graphene lattice. When Gd is located at the edge of nanoribbon ($h_1$), there is a spin imbalance of the two edge states and an asymmetric edge states consists of both spin-up and spin-down components appears. This is because of $d$-orbitals with the angular quantum number $l$ ($l$ = 2) of Gd no longer maintaining center symmetric attributed to a transverse electric field from left to right induced by the off-center asymmetry adsorption of Gd on ZGNR(Gd is adsorbed off the center of carbon hexagon by 0.026 Å in $h_1$ site), and the spin polarizations at each edge site are remaining opposite due to the topology of graphene lattice. While Gd locates at the center of nanoribbon ($h_2$), the spin density is symmetric in ZGNR plane and the FM behavior consists of spin-down states at two edge states appears. In this case, the original center symmetric of $d$-orbitals of Gd is not broken due to the central symmetry adsorption of Gd (Gd adatom is right at the center of carbon hexagon), which would polarize the spin directions at two nearest neighbor C atoms of Gd from antiparallel to parallel, and consequently lead to the FM states at two edges of ZGNR plane. Therefore, we suggest that the different hybridization between C-$p_z$-up/down states at two edges and Gd-$sd$ states in $h_1$/$h_2$-Gd-ZGNR is associated with the different edge states affected by adsorption location of Gd atom, which would essentially contribute to the distinction of electronic structures. On the other hand, the second-order MAE determined by the coupling between unoccupied and occupied states around Fermi level in electronic structures of $h_1$/$h_2$-Gd-ZGNR would behave different, as we will discussed in Sec. III. C3.

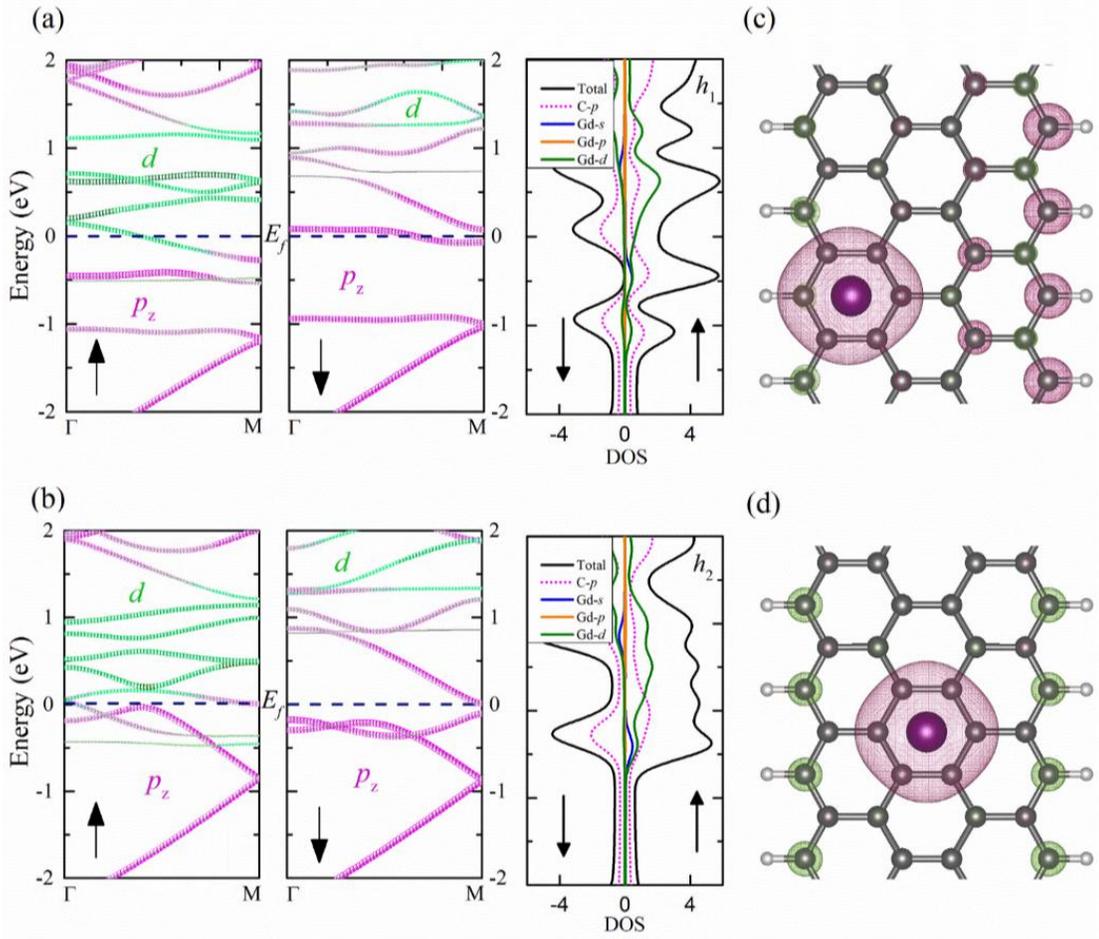

FIG. 2. (Color online) Orbital projected electronic band structures and density of states (DOS) of Gd-4ZGNR with Gd adsorbed in (a) $h_1$ and (b) $h_2$ sites. Spin-up (left)/down (right) bands showing the amplitudes of the projection band of the carbon $p_z$ contributions (red region) and Gd-5$d$ contributions (green region). The Fermi energy is set as zero. The $s$, $p$, $d$ orbitals distribution are marked by different colors in the corresponding DOS plots. Spin density plots ($3 \times 10^{-3}$ $e$/Å$^3$) of Gd-4ZGNR matrix for (c) $h_1$ and (d) $h_2$ cases. The pink gray densities correspond to spin-up and green gray to spin-down components.

Furthermore, we explain the origin of magnetic moment for $h_1/h_2$-Gd-4ZGNR based on above electronic structures. Because of that the partially filled orbitals always play an active role in the formation of magnetic moment, thus it can be inferred that the contribution to magnetic moment excluding Gd-4$f$ in Gd-ZGNR on Gd adatom mainly comes from $5d_{x^2-y^2}$, $d_{z^2}$ orbitals, with the contribution of $s$-orbital in small amount, as revealed by the partial DOS of Fig. 2(a) and (b). As shown in the spin-density of Fig. 2(c) and 2(d), it can be seen that the carbon atoms in ZGNR planes have an obvious positive/negative contribution (mainly comes from $p_z$ orbitals) to the total magnetic moment. For the $h_1$-Gd-4ZGNR, the spin polarizations at two edge are anti-aligned each other and the component of spin-up states is more than that of spin-down states in ZGNR plane, due to the off-center asymmetry adsorption of Gd. While for the $h_2$-Gd-4ZGNR, the spin density in ZGNR plane maintain bilateral symmetry but turns to FM behavior at two edges due to the central adsorption symmetry of Gd, where the spin-down density plays a dominant role. The distribution of spin density for $h_1/h_2$-Gd-4ZGNR provides an intuitively understanding for the considerable different magnetic moment at different adsorption site. Hence, the total magnetic moment with excluding the $f$ orbital of Gd-ZGNR at both sites (2.20 $\mu_B$ in $h_1$ and 1.07 $\mu_B$ in $h_2$) is determined by the partially filled orbitals ($5d_{x^2-y^2}$, $d_{z^2}$ and 6$s$) of Gd and different spin polarization of C-$p_z$ orbitals in ZGNR plane.

### C. Magnetic anisotropy in Gd-ZGNR

#### 1. Angle-dependent MAE and MAE contributions in *k*-space

Since the SOC and the intra-atomic exchange are of the same magnitude in Gd element, we perform further calculations to investigate the magnetic anisotropy of Gd-ZGNR based on Force theorem [40]. In general, the MAE presents the most fundamental quantities of magnetic systems as its sign identifies the easy magnetization axis and its magnitude provides an estimate of the stability of magnetization in the application of spintronic data storage and processing. Firstly, we calculate the angular ($\theta$) dependent of MAE at 0º≤ $\theta$ ≤180º and fit the angle-dependent MAE on cos$\theta$ for the $h_1$/$h_2$-Gd-ZGNR system in Fig. 3(a) and 3(b), where angle $\theta$ is defined between the quantization axis of spin and the vertical *z*-axis. The polynomial model fitting for the angle-dependent MAE is used for estimating the parameters of perturbation, where the well-fitted coefficient of cos$\theta$ and sin$^2\theta$ indicates the contribution from the first-order perturbation and second-order perturbation in Gd-ZGNR. Our fitting parameters for $h_1$/$h_2$-Gd-4ZGNR are provided in the inset of Fig. 3, which generally reveals the MAE originates from the second-order perturbation [47]. Actually, the MAE originates from the competition between parallel and perpendicular contributions of the SOC. The value of MAE is defined as the total energy difference between the magnetic states with spin moments aligned in and out of plane( $\Delta E = E_{(100)} - E_{(001)}$ ). Here, the values of MAE for $h_1$/$h_2$-Gd-4ZGNR are calculated as 3.44 and 3.92 meV, respectively, indicating that the out-of-plane magnetization is preferred for both cases.

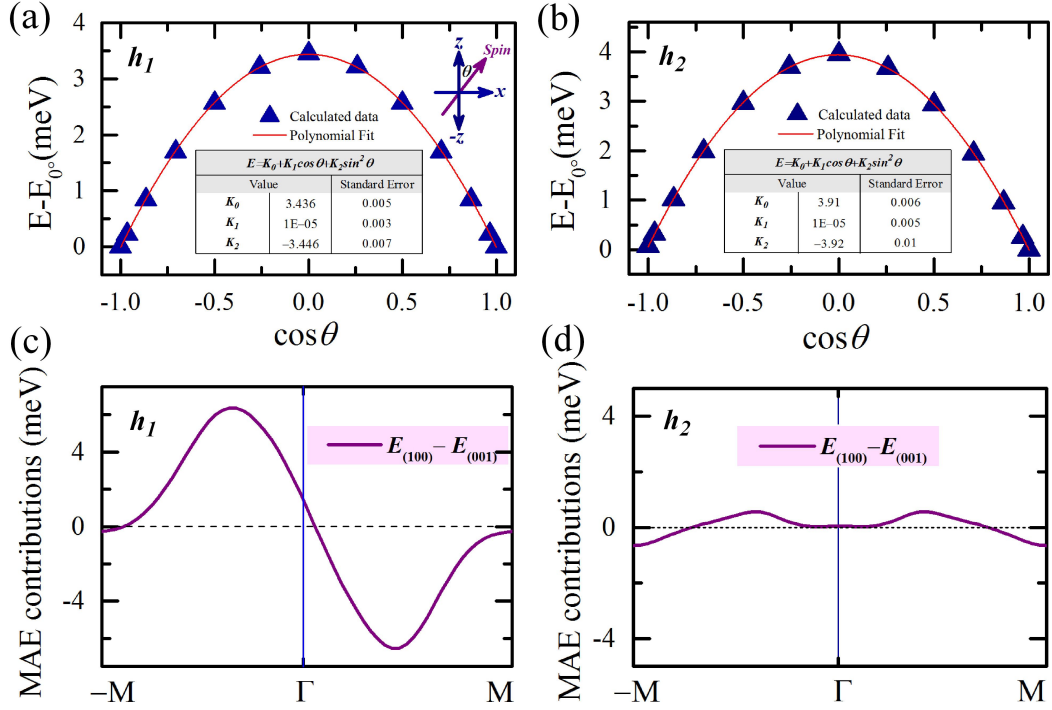

FIG. 3. (Color online) The polynomial fitting of angle($\theta$)-dependent MAE at $0° \leq \theta \leq 180°$ for (a) $h_1$ and (b) $h_2$-Gd-4ZGNR. The energy for the case of the spin along the $z$ direction is set to zero. The inset of (a) illustrates the rotation of spin vector $S$ on the $-z\sim x\sim z$ plane with the angle $\theta$. The polynomial fitting formula, parameters and standard errors are also presented. The MAE contributions calculated by Eq. (1) in FBZ ($-M$-$\Gamma$-M) for (c) $h_1$ and (d) $h_2$-Gd-4ZGNR cases.

Furthermore, we examine the MAE contributions in $k$-space for above two cases, which has been well-established for other magnetic systems [47-49]. It is defined by

$$\Delta E = E_a - E_b = \sum_{k \in FBZ} \left( \sum_{i \in o} \varepsilon_i^a(k) - \sum_{i \in o'} \varepsilon_i^b(k) \right) \qquad (2)$$

where the $\varepsilon_i^a(k)$ and $\varepsilon_i^a(k)$ is the energy of the $i^{th}$ band at point $k$ in the first Brillouin zone (FBZ) for the magnetic states with spin moments aligned on the direction of $a$ and $b$. The $o$ and $o'$ specify the occupied states for the $a$ and $b$ direction, respectively. By using Eq. (2), the MAE contributions in *k*-space for $h_1/h_2$-Gd-4ZGNR are calculated, respectively, as shown in Fig. 3(c) and 3(d). It is interesting to note that, the MAE exhibits a novel counterbalanced trend between −M-Γ and Γ-M in FBZ for $h_1$-Gd-4ZGNR (Fig. 3(c)), which cannot be traditionally derived from the second-order perturbation of spin-orbital coupling in the form of $\lambda \mathbf{L} \cdot \mathbf{S}$. Whereas for the $h_2$ case, Γ-centered symmetrical MAE is found in Fig. 3(d), consistent with the trend of traditional second-order perturbation. We suppose that the counterbalanced MAE contributions in *k*-space of $h_1$-Gd-4ZGNR comes from the hidden first-order perturbation.

### 2. First-order MAE and Rashba effect

In this section, the unconventional magnetic MAE of 1D Rashba system is further elaborated by reproducing the MAE distribution in *k*-space from the Rashba spin splitting. To explore the origin of above counterbalanced MAE in $h_1$-Gd-ZGNR, a direct comparison of band structures with SOC-perturbed in different magnetization directions is provided in Fig. 4, where one can see an obvious movement around the Γ point in the band structures with the $(001)$ and $(00\bar{1})$ magnetization direction (Fig. 4($a_2$) and 4($a_3$)). To quantify the MAE contribution of some specific bands, we calculate the occupied states changes between $(00\bar{1})$ and $(001)$ magnetization for one of the bands by rewriting Eq. (2) explicitly as:

$$\Delta E' = E_{-z}' - E_z' = \sum_{k \in FBZ} \left[ \varepsilon_{-z}(k) f_{-z}(k) - \varepsilon_z(k) f_z(k) \right] \tag{3}$$

where the $f_{-z}(k)$ and $f_z(k)$ represents the occupation probability of one band for the direction of $(00\bar{1})$ and $(001)$, respectively. It is noticed that, the $E_{(00\bar{1})} - E_{(001)}$ in *k*-space calculated by Eqs. (2) and (3) can be considered twice the first-order perturbation contribution. According to Eq. (3), the $\Delta E'$ of '*B1*', '*B2*' bands and their combination is found as the major part of first-order MAE along −M-Γ-M line, as plotted in Fig. 4($b_1$), ($b_2$) and ($b_3$), respectively. As a comparison, the total first-order MAE contribution in FBZ is also provided in Fig. 4($b_3$). It is noted that the first-order part has zero net contribution to the total MAE, since it is an odd function in *k*-space. There is no doubt that, the opposite movement of band structure around Γ point of $h_1$-Gd-4ZGNR in the (001) and $(00\bar{1})$ magnetization direction indicates a Rashba-type spin splitting, in the presence of strong SOC strength and broken inversion symmetry. Because of SOC, the resulting dispersion of the spin-split bands can be described by $E^{\pm}(k)=\frac{\hbar^2}{2m^*}k_x^2 \pm \alpha k_x$, where $\alpha$ is the Rashba parameter, $k_x$ the in-plane momentum, $m^*$ the electron effective mass. For the present $h_1$-Gd-4ZGNR, the moment splitting $\Delta k_R$ of the Rashba bands from the crossing point is 0.017 Å$^{-1}$, as defined in Fig. 4(d). Using the calculated $\Delta k_R$ and the fitted effective mass ($m^* = -0.067m_0$), the Rashba parameter $\alpha = \hbar^2 D k_R/m^*$ (see the derivation in Supplemental Material) is deduced to 1.89 eV Å, which is comparable to that of 1D Pt nanowire as a giant Rashba system [15]. Actually, the effective SOC-splitting strength between bands '*B1*' and '*B3*' as the spin-axis along the vertical direction could be derived from the band shift. The first-order perturbation enlarges the non-perturbed gap (Δ) as $\Delta_\theta = \Delta\sqrt{1+4\left(\frac{h}{\Delta}\right)^2}$ (Here $h = \xi cos\theta$, where $\xi$ is the effective SOC interaction constant, the detail of Pseudo-gap model is shown in the Supplemental Material). Hence, the pseudo-gap $\Delta_\theta$ is proportional to $(cos\theta)^2$ when the polar angle $\theta$ is close to 90º, and proportional to $\sim cos\theta$ when $\theta$ is close to 0º. Above model could apply to the spin-axis in (*x*,-*z*) plane as well. Whereafter, we calculate the $\Delta_\theta$ with varying the angle of spin-axis in 0º ≤ $\theta$ ≤ 180º as shown in Fig. 4(c), which is well consistent with our Pseudo-gap model. According to our calculated results, the gap at the Γ point enlarged by SOC splitting as $\Delta_\theta$=131.35 meV in the (001) magnetization. Then, we can directly figure out the maximum effective SOC strength $\xi \approx$ 65.6 meV ($\Delta_\theta \approx 2h = 2\xi cos\theta$, $\theta = 0º$) in $h_1$-Gd-4ZGNR, which is close to the estimated SOC coupling constant (60 meV) in Gd adsorbed freestanding graphene [44]. Such strong SOC strength ($\xi \approx$ 65.6 meV) exist in the giant 1D Rashba system ($h_1$-Gd-4ZGNR) is reasonable. Based on the discussions above, we reproduce the *k*-space distribution of the first-order MAE from two specific bands showing the Rashba spin-split attributed to strong SOC (~ 65.6 meV) and broken inversion symmetry, which definitely confirms that the first-order MAE is from the Rashba effect. This approach quantitatively proves that the Rashba spin-split has significantly effect on the MAE contributions in *k*-space, which will be inevitably adopted in future works focusing on magnetic anisotropy in other Rashba systems.

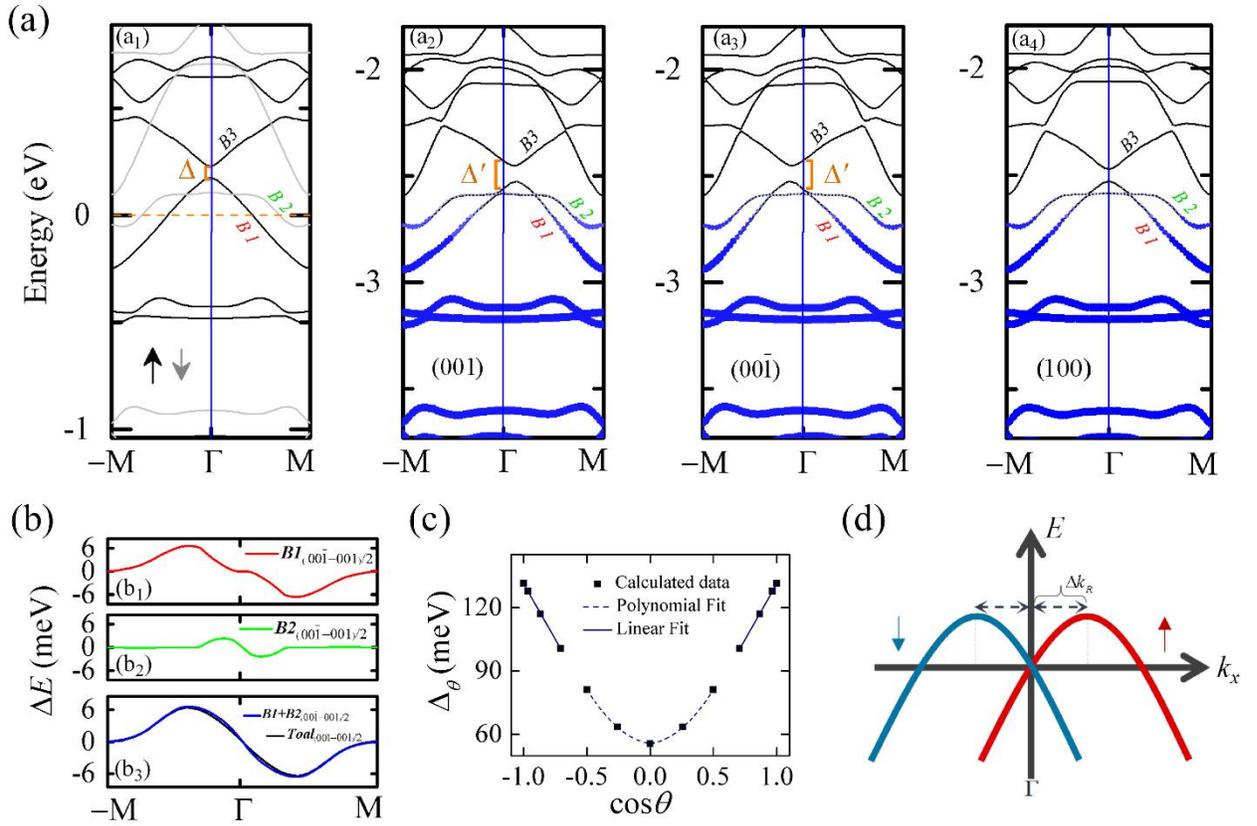

FIG. 4. (Color online) The counterbalanced first-order MAE for $h_1$-Gd-4ZGNR. (a) The band structures of ($a_1$) non-perturbed and perturbed with ($a_2$) (001), ($a_3$) (00$\bar{1}$) and ($a_4$) (100) magnetization in FBZ. The Fermi energy is set as zero in ($a_1$). The occupation probabilities of electrons are visualized by the relative size of blue bubble (0~1) in ($a_2$), ($a_3$) and ($a_4$). The bands '*B1*', '*B2*' and '*B3*' are marked. (b) The first-order MAE contributions of band ($b_1$) '*B1*', ($b_2$) '*B2*' and ($b_3$) their summation along the $-$M-Γ-M line. For comparison, the total first-order MAE contribution in FBZ is also plotted in ($b_3$). (c) The dependence of the pseudo-gap ($\Delta_\theta$) at Γ point on $\cos\theta$. (d) Red and blue

parabolic lines around the Γ point are interpreted as Rashba-type up/down-spin splitting bands. The $\Delta k_R$ is calculated as 0.017 Å in $h_1$-Gd-4ZGNR.

We also calculate the band structures with different magnetization directions of $h_2$-Gd-4ZGNR (Fig. S3 of Supplemental Material) to explore whether the band shift exist or not at the metastable central-hollow $h_2$ site. Obviously, there is no shift in band structures perturbed with SOC, which is strongly expected because of that the MAE contribution is strictly Γ-point centered symmetrical in FBZ (Fig. 3(d)). It is believed that the first-order MAE as well as Rashba effect of Gd-4ZGNR relies on the structural symmetry based on Gd adsorption site. There is an obvious Rashba spin split in the SOC enabled band structure of $h_1$-Gd-4ZGNR, where the Gd atom is adsorbed off the center of nanoribbon by 0.026 Å. However, no Rashba spin split in the $h_2$-Gd-4ZGNR, where the Gd adatom is right at the center of nanoribbon. This discrepancy is attributed to the different symmetries of $h_1$ and $h_2$-Gd-4ZGNR, of which the former implies an electric field transverse to the nanoribbon and consequently the Rashba spin split if the spin axis has a finite component perpendicular to the nanoribbon. The Rashba effect predicted in $h_1$-Gd-4ZGNR is therefore due to the electric field *transverse* to the nanoribbon associated with the off-center adsorption of Gd atom on $h_1$ site.

In conclusion, we prove that the Rashba spin split may significantly impact the MAE contributions in *k*-space, where the horizontal band shift of band structures caused by Rashba effect could lead to a significant change of occupied states near $E_f$ along *k*-line, and consequently the counterbalanced first-order MAE contributions in FBZ. Although the net contribution from Rashba effect to the value of total MAE is zero, the approach presented here may help to design materials with a finite net contribution.

### 3. Second-order MAE

As we know, the second order perturbation of MAE can be related with the changes in the electronic structure around the Fermi level by using the approach proposed by Wang, Wu, and Freeman [47]:

$$\Delta E \sim \sum_{u,o} \frac{\left| \langle u | l_z | o \rangle^2 - \langle u | l_x | o \rangle^2 \right|}{\varepsilon_u - \varepsilon_o} \quad (4)$$

where $u$ and $o$ specify the unoccupied and occupied minority spin states, respectively, and the $l_x$ and $l_z$ are angular momentum operators. The pair of unoccupied and occupied states around Fermi level is important to the second-order MAE contribution. Similar analyses were applied to investigate the magnetic anisotropy of other magnetic systems [40,50].

To further quantify understand the real second-order perturbation in $h_1$-Gd-ZGNR, here, we subtract the known first-order perturbation contribution (Fig. 4(b$_3$)) from the total MAE ($E_{(100)} - E_{(001)}$) in FBZ (Fig. 3(b)) through Eq.(2). The subtraction result can be thought of as the pure second-order MAE contributions, as shown in Fig. 5(a). On the other hand, the second-order perturbation of $h_2$-Gd-4ZGNR is proved to be the only source of MAE (Fig. 5(b)). In general, the couplings among up-spin electronic states ($\Delta E_{\uparrow\uparrow}$), down-spin states ($\Delta E_{\downarrow\downarrow}$), up and down spin states ($\Delta E_{\uparrow\downarrow+\downarrow\uparrow}$) might make certain contributions to the second-order MAE. For $h_1$/$h_2$-Gd-4ZGNRs, $\Delta E_{\downarrow\downarrow}$ only contributes a few to MAE near M point, the major MAE contribution in $k$-space comes from $\Delta E_{\uparrow\uparrow}$. As shown in Figs. 5(c) and 5(d), we marked the orbital hybridizations between Gd-$sd$ orbitals of up/down-spin bands with different colors.

By analyzing the up-spin band dispersion of $h_1$-Gd-4ZGNRs, one can clearly see that the $d_{x^2-y^2}$ states across $E_f$, $d_{xz}$ and $d_{z^2}$ are distributed to the region not only above but also below the $E_f$, both contain the occupied and unoccupied states. While the $d_{xy}$ is nearly empty ($d_{yz}$ is lies higher than $d_{xy}$ in conduction band and thus the contribution to MAE is negligible). Note that, the magnetic quantum number of $d_{z^2}$, $d_{xz}/d_{yz}$, $d_{x^2-y^2}/d_{xy}$ is 0, ±1, ±2, respectively. Therefore, the positive MAE value at the front end of $k$-line (~2/5 Γ-M) can be explained by the perturbation interaction of near-degenerate $d_{x^2-y^2}$ ($d_{xy}$) states, and the coupling between the unoccupied $d_{xy}$ states and small amounts of occupied $d_{x^2-y^2}$ states (the occupation probability is less than 0.5) through the $l_z$ operator; Whereas at the second half (~3/5 Γ-M), the coupling of $\langle d_{xz}(d_{x^2-y^2})|l_x|d_{x^2-y^2}(d_{xz})\rangle$ contributed negatively to MAE from −0.3 eV to 0.6 eV (the coupling of unoccupied $d_{z^2}$ with occupied $d_{x^2-y^2}$ states is zero), corresponding to the second part of Eq. (4). In the $\Delta E_{\downarrow\downarrow}$, some negative MAE contribution attributed to the coupling between some occupied $d_{x^2-y^2}$ and unoccupied $d_{xz}$ states in −0.1 eV ~ 0.1 eV near M point.

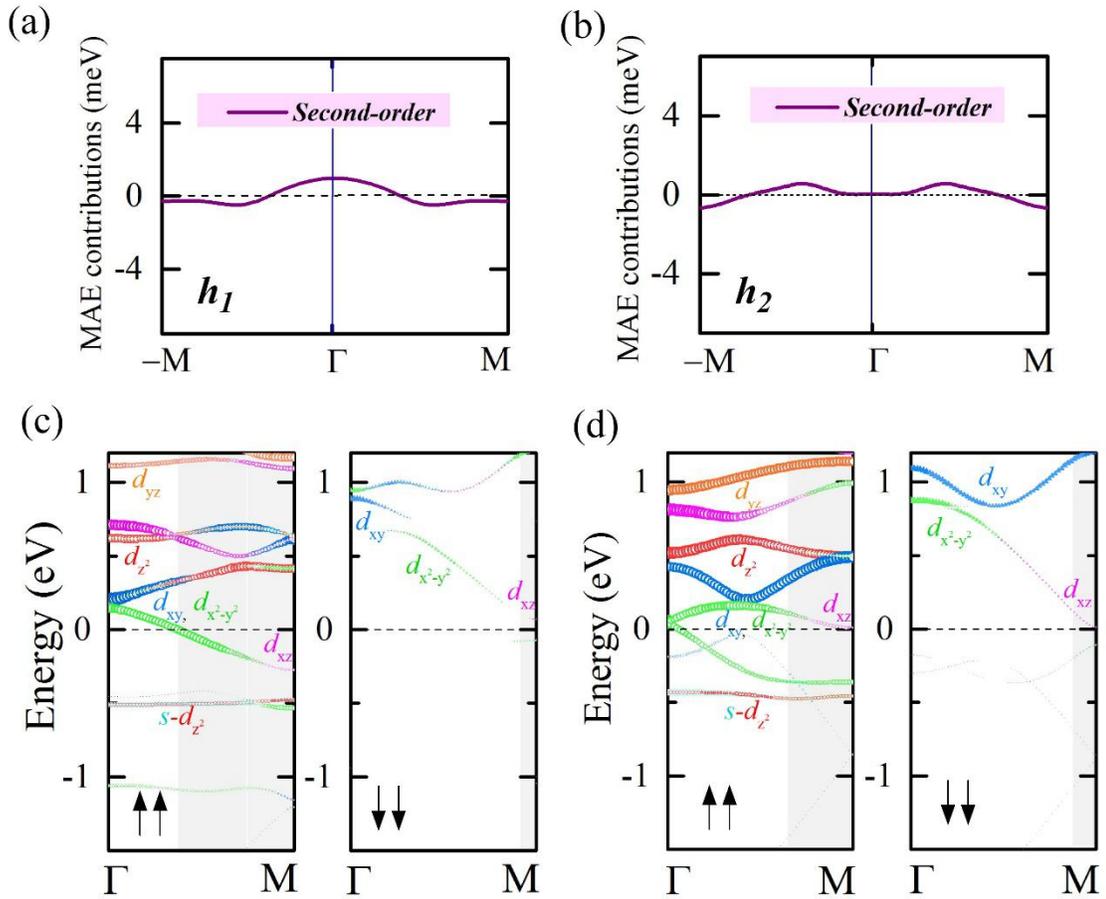

FIG. 5. (Color online) The real second-order MAE contributions for (a) $h_1$ and (b) $h_2$-Gd-4ZGNRs. The up/down-spin band dispersions of Gd 6$s$-5$d$ orbitals for (c) $h_1$ and (d) $h_2$-Gd-4ZGNRs. The blue, green, rose red, orange and red bands highlight Gd-$d_{xy}$, $d_{x^2-y^2}$, $d_{xz}$, $d_{yz}$ and $d_{z^2}$ contributions in spin-up/down channels, respectively. The $E_f$ is set as zero.

The MAE of metastable $h_2$-Gd-4ZGNR (Fig. 5(b)) could be directly explained by the second-order perturbation theory combined with orbital-projected band structures in Fig. 5(d). It is obviously that $d_{xy}$, $d_{x^2-y^2}$ and $d_{z^2}$ states in the up-spin channel have considerable contributions above and below the $E_f$. In the up/down-spin channel, a few component of $d_{xz}$ exist near $E_f$, which should be considered as thecontributory unoccupied states. Besides, $d_{yz}$ orbital in a higher energy level contributes to MAE is negligible). Thus, the positive MAE value at the front end of $k$-line (~4/5 Γ-M) is contributed by the $\langle d_{xy}(d_{x^2-y^2})|l_z|d_{x^2-y^2}(d_{xy})\rangle$ from −0.4 eV to 0.5 eV through the $l_z$ operator (note that the greater positive value at ~2/5 Γ-M is contributed by the near-degenerate $d_{x^2-y^2}$ ($d_{xy}$) states interaction); Whereas at the second half (~1/5 Γ-M), the coupling of $\langle d_{xz}|l_x|d_{x^2-y^2}\rangle$ at −0.4 eV ~ 0.02 eV contributed negatively to MAE. In addition, the coupling between occupied $d_{x^2-y^2}$ and unoccupied $d_{xz}$ states of $\Delta E_{\downarrow\downarrow}$ also has some contribution to the negative MAE around M point (in the energy region of −0.1 eV to 0.02 eV).

### D. Gd adsorbed ZGNRs with different widths

Due to the width of graphene nanoribbon could reaches a few nanometer in practical experiment, we also study whether the Rashba effect stabilizes or not in a large Gd-ZGNR system. Therefore, the adsorption stability and MAE behavior in $k$-space of Gd-ZGNRs with various widths ($n$=5, 6, 7, 8, 9, 10, 12) are further examined. The calculated magnetic ground states for pure $n$ZGNRs ($n$=5, 6, 7, 8, 9, 10, 12) show that the AFM state is always the most favorable one (see Table S1 in the Supplemental Material), which are same as that of 4-ZGNRs. The width of 12-ZGNR reaches approximation to 2.4 nm, and the Gd-12ZGNR system corresponds to a coverage of one adatom per 96 carbon atoms. The adsorption concentration of Gd-$n$ZGNRs ($n$=5, 6, 7, 8, 9, 10, 12) correspond to 2.5%, 2.08%, 1.79%, 1.56%, 1.39%, 1.25%, 1.04%, respectively. By calculating the binding energies of Gd-$n$ZGNRs at various possible sites (the inequivalence adsorption sites are labeled in Fig. S4) using Eq. (1), it can be concluded that the edge hollow site ($h_1$) is also energetically favorable for all examined Gd-$n$ZGNRs, as shown in Fig. 6(a). By using the Force theorem, the values of MAE for various $h_1$-Gd-$n$ZGNRs ($n$=4, 5, 6, 7, 8, 9, 10, 12) systems are calculated as 3.44, 3.799, 3.915, 3.210, 2.713, 3.438, 3.806, 3.78 meV, respectively. Perpendicular MAE is found in all cases, and the MAE values of Gd-$n$ZGNRs show no obvious dependency on the coverage of Gd atom.

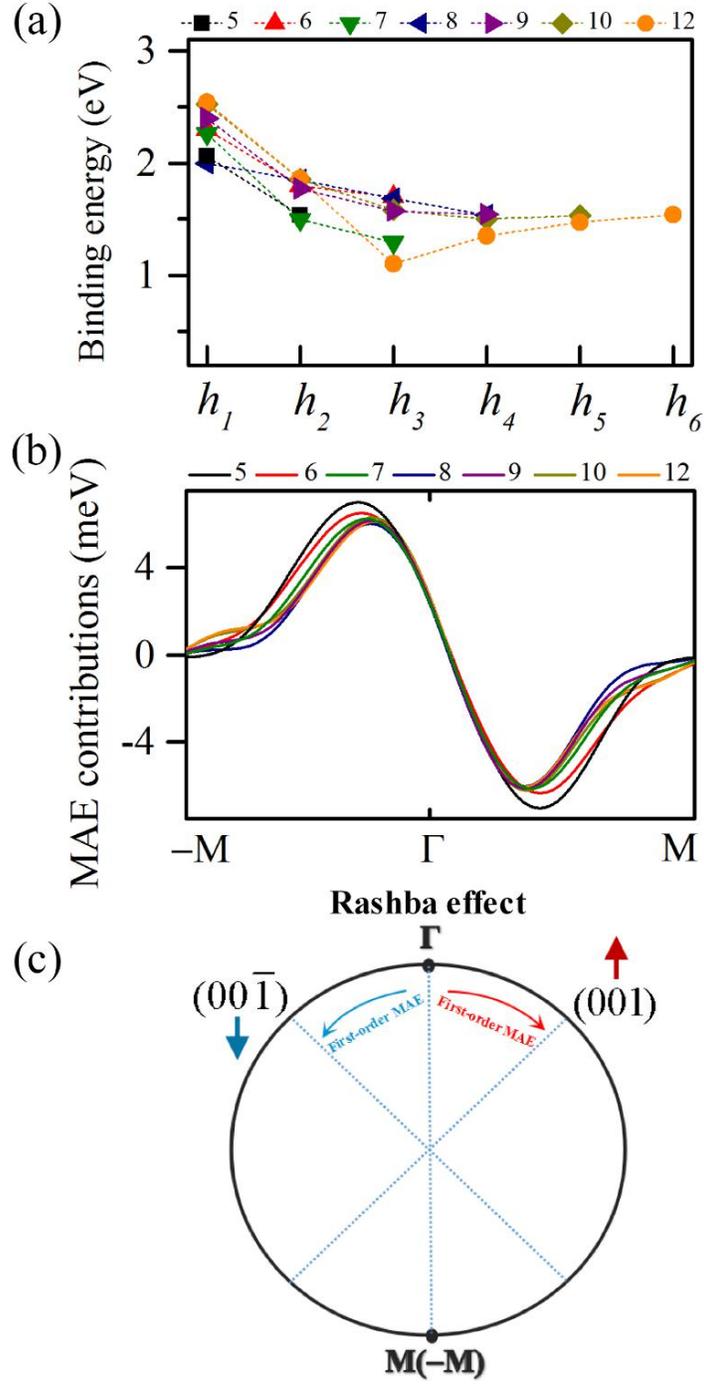

FIG. 6. (Color online) (a) Binding energies of Gd-$n$ZGNRs ($n$=5, 6, 7, 8, 9, 10, 12) at possible adsorption sites. The abscissa consists of the possible adsorption hollow sites. (b) The MAE contributions in FBZ for the preferred $h_1$-Gd-$n$ZGNRs. (c) The schematic diagram for the first-order MAE attributed to Rashba effect in Gd-ZGNR.

Then, the MAE contributions in FBZ calculated by Eq. (2) for preferred $h_1$-Gd-$n$ZGNRs are provided in Fig. 6(b), all of which show similar trends as that of $h_1$-Gd-4ZGNR (Fig. 3(c)). The Γ-centered counterbalanced trend of MAE contributions for $h_1$-Gd-$n$ZGNRs further validates the robust Rashba effect in wider $h_1$-Gd-$n$ZGNRs systems according to the established microscopic connection between first-order MAE and Rashba effect in Sec. III C2, no matter whether the width is even or odd number. It is also noted that, the amplitude of first-order MAE contributions are roughly the same for various Gd-$n$ZGNRs, which suggests that Rashba effect as well as first-order MAE of Gd-ZGNR has no obvious dependency on the coverage of Gd atom. In particular, the Rashba effect can manipulate the electron trajectory determines by the spin direction in $k$-space, and consequently creates a counterbalanced first-order MAE in FBZ for various Gd-$n$ZGNRs, as revealed in Fig. 6(c). Only thing to note is that Gd adsorbed on armchair graphene nanoribbons are not considered here, which may show diversified phenomenon due to different edge states and symmetry. Finally, we would like to suggest some useful rules which will guide the future exploration of other 1D Rashba sysems.  As discussed above, the intrinsic nature of Rashba effect is determined essentially by the electric field *transverse* to the nanoribbon associated with the off-center adsorption of Gd atom. In general, the asymmetric hollow site at nanoribbons edge is the most stable one for 3$d$ elements V or Co (maybe other elements remain to be studied) adsorbed on ZGNRs, which would increase the possibility of realizing Rashba effect in one-dimensional system. Therefore, the Rashba effect may also appear spontaneously in other ZGNR-based systems where elements with large spin-orbit coupling prefer to adsorb on the asymmetric hollow site, which has been ignored by previous studies on foreign atom adsorbed nanoribbon. We look forward to the future theoretical/experimental studies on realizing the Rashba spin-split in ZGNR-based systems.

## IV. CONCLUSIONS

In summary, we have carried out a comprehensive study on the electronic and magnetic properties of Gd adsorbed zigzag graphene nanoribbons using the first-principles methods. It is found that the orbital hybridization, magnetic moment and magnetic anisotropy in *k*-space of Gd-ZGNR rely on the Gd adsorption sites (edge $h_1$ and center $h_2$). Via analysis of the hybridization between Gd-*sd* and C-*p_z* orbitals in electronic structure, the origin of magnetism in $h_1$/$h_2$-Gd-ZGNR is clarified. The angle-dependent MAE is calculated for $h_1$/$h_2$-Gd-ZGNR, which both reveal a second-order perturbation. It should however be noted that the unconventional MAE contribution in k-space has been discovered in the self-assembled $h_1$-Gd-ZGNR besides the conventional second-order MAE, while it is absent in the $h_2$-Gd-ZGNR. The unconventional MAE is further elaborated by reproducing the MAE distribution in *k*-space from the Rashba spin splitting (1.89 eV Å) in Gd-ZGNR attributed to the strong SOC strength (~65.6 meV) and broken inversion symmetry, and definitely confirms the microscopic connection between first-order MAE and Rashba effect. Moreover, the real second-order MAE contribution of Gd-ZGNR is explained by the second-order perturbation theory in detail. The first-order MAE as well as the Rashba spin splitting is robust in the Gd-ZGNR systems with various widths. Our work opens a new pathway in looking for 1D Rashba spin-split systems and provides deep understanding of the connection between Rashba nature and unconventional magnetic anisotropy.

ACKNOWLEDGMENT

This work was sponsored by the National Basic Research Program of China (Grant No. 2011CB606405), the CAEP Microsystem and THz Science and Technology Foundation (Grant No. CAEPMT201501), and the Science Challenge Project (No. TZ2016003).

# Supplemental Material

# Unconventional magnetic anisotropy in one-dimensional Rashba system realized by adsorbing Gd atom on zigzag graphene nanoribbons

Zhenzhen Qin[1], Guangzhao Qin[2], Bin Shao[3], Xu Zuo[1*]

[1] *College of Electronic Information and Optical Engineering,*

*Nankai University, Tianjin, 300350, China*

[2] *Institute of Mineral Engineering, Division of Materials Science and Engineering,*

*Faculty of Georesources and Materials Engineering,*

*RWTH Aachen University, Aachen 52064, Germany*

[3] *Bremen Center for Computational Material Science,*

*Bremen University, Am Fallturm 1, Bremen 28359, Germany*

Email: xzuo@nankai.edu.cn

---

[*] To whom correspondence should be addressed.

## (a) Electronic structure of pure zizag graphene nanoribbon

Fig. S1 gives the structure, sin density and band structures of pure 4ZGNR. From Fig. S1(b), it is clearly see that spin moments are mainly distributed at the edge carbon atoms and slower decay towards the center of the ribbon. The magnetic moment fluctuation across the ribbon arises from quantum interference effects caused by nanoribbon edges. Due to topology of the lattice, the atoms of the two edges belong to different sublattices of the bipartite graphene lattice. The spin density on the C atoms on one edge are antialigned to the spin density on the opposite edge, and the polarizations of neighboring sites belonging to different sublattices are also opposite. The band structure of ZGNR decorated with up/down C-$p_z$ orbitals is presented in Fig. S1(c) respectively.

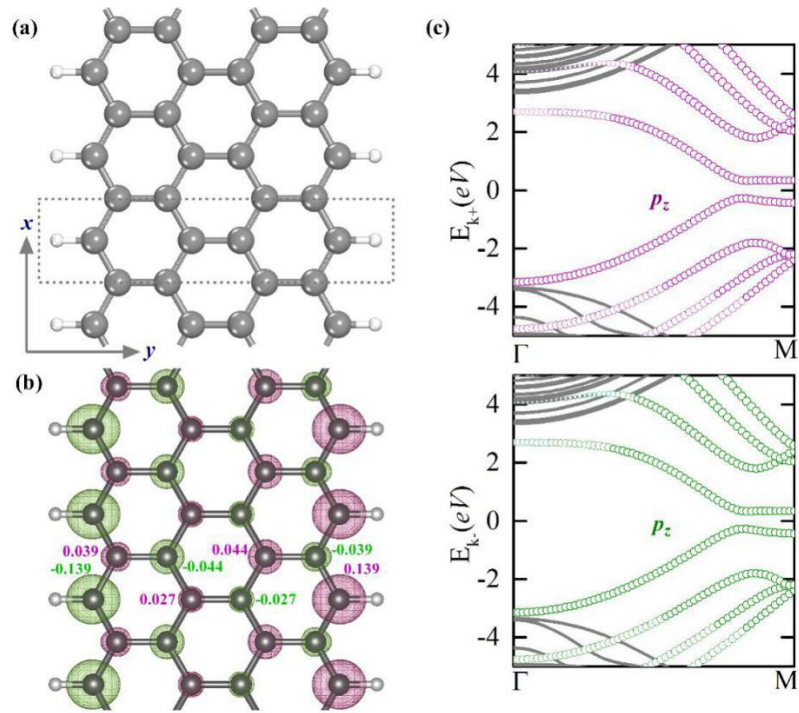

Figure S1. Structure of H-passivated 4-ZGNR, where the gray and white balls represent C and H atoms, respectively. The unit cells are indicated by dashed lines. (b) Spin up (red gray) and down (green gray) density plots of pure 4-ZGNR ($2\times10^{-3}$e/Å$^3$) showing together with the corresponding values of local magnetic moments. (c) Spin up (red gray)/down (green gray) bands of 4ZGNR showing the amplitudes of the projection band of $p_z$ orbital of edge carbon atoms.

## (b) Orbital-projected density of states of Gd-ZGNR

We provide the orbital-projected density of states of Gd-ZGNR at $h_1$ site, and give that of $h_2$ site for comparison, as shown in Fig. S2. It is found that the orbital splitting of Gd-5$d$ levels are almost no difference in the two adsorbed cases: The down-spin density of Gd-$d$ orbitals are all empty, while the up-spin density reveals that $d_{xy}$, $d_{yz}$, $d_{xz}$ orbitals are nearly unoccupied, which lie higher in energy than the obvious partially filled $dx^2$-$y^2$ and $dz^2$ orbitals.

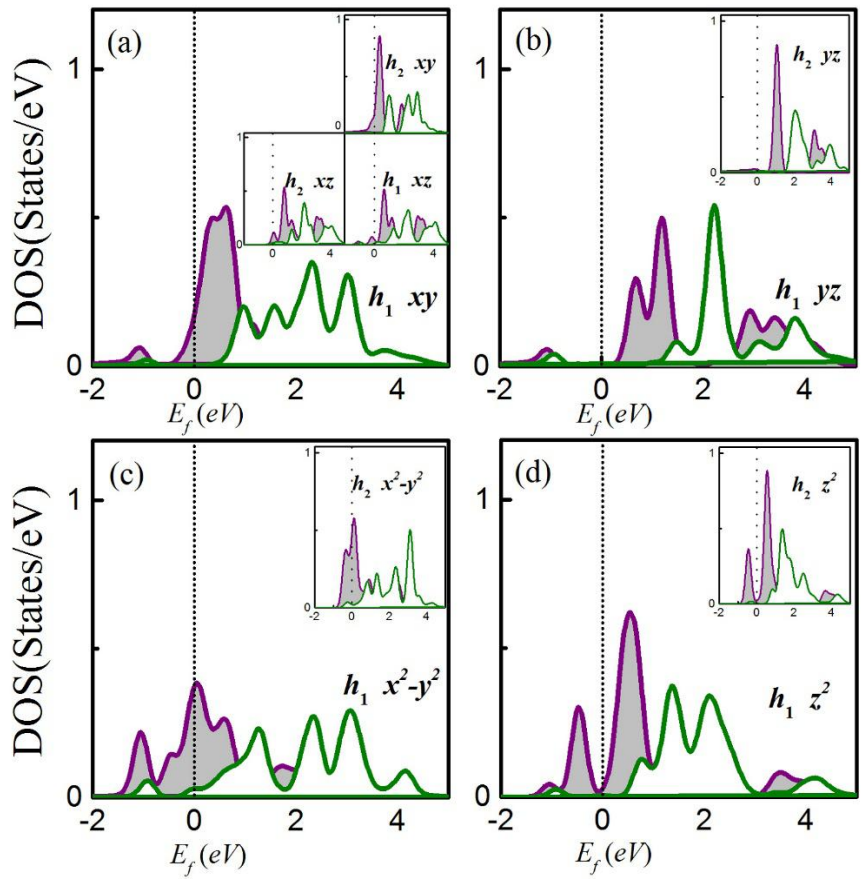

Figure S2. The orbital-projected densities of states of Gd adatom at $h_1$ position in ZGNR. Insets of (a-d) present $d$-DOS of Gd adatom at $h_2$ site. And insets of (a) provides the $d_{xz}$ partial DOS of Gd adatom at $h_1$ site. Red gray region denotes spin-up contributions and green blank the spin-down contributions.

## (c) Pseudo gap in $h_1$-Gd-ZGNR

Due to the pseudo-gap is enlarged at $\Gamma$ point in the perturbed band structures of $h_1$-Gd-ZGNRs, we prove that relationship from the first-order perturbation of the SOC interaction, $H_{SOC} = \xi l \cdot s = \xi(l_x \sin\theta + l_z \cos\theta)s$, where we assume that the spin is in the $xz$-plane, here $s = 1/2$, $\xi$ is the effective SOC interaction constant, and $\theta$ is the angle between the spin-axis and the $z$-axis. Because the two separated orbitals at $\Gamma$ point are $d_{xy}$ and $d_{x^2-y^2}$, so we can directly write the matrix representation

$$H_{SOC} = \frac{\xi}{2}\cos\theta \begin{pmatrix} \varepsilon & -2i \\ 2i & \varepsilon+\Delta \end{pmatrix}, \quad (1)$$

where $\varepsilon$ and $\varepsilon+\Delta$ are the two levels before perturbation at $\Gamma$ point.

Set the $h = \xi\cos\theta$, it can be written as

$$\begin{pmatrix} \varepsilon & -ih \\ ih & \varepsilon+\Delta \end{pmatrix}, \quad (2)$$

Diagonalizing this matrix, we obtain two levels after perturbation

$$\lambda_\pm = \frac{2\varepsilon+\Delta \pm \sqrt{\Delta^2+4h^2}}{2}, \quad (3)$$

and the gap after perturbation

$$\Delta_\theta = \sqrt{\Delta^2+4h^2} = \Delta\sqrt{1+4\left(\frac{h}{\Delta}\right)^2}. \quad (4)$$

When $h \ll \Delta$, $\Delta_\theta \approx \Delta\left(1+2\left(\frac{h}{\Delta}\right)^2\right)$, and when $h \gg \Delta$, $\Delta_\theta \approx 2h$. Note that $h \propto \cos\theta$, when the polar angle $\theta$ is close to 90, the gap is proportional to $(\cos\theta)^2$, and when $\theta \sim 0°$, the gap is proportional to $\cos\theta$.

### (d) Band structures of $h_2$-Gd-ZGNR perturbed with SOC

The band structures for $h_2$-Gd-4ZGNR in different magnetization directions is provided in Fig. S3. Observed from the band structures with the perpendicular (001) and (00$\bar{1}$) magnetization, it is clearly shown that there is no movement which turn out that the non-existed offsetting first-order perturbation contribution, consistent with the MAE contributions along $k$-line.

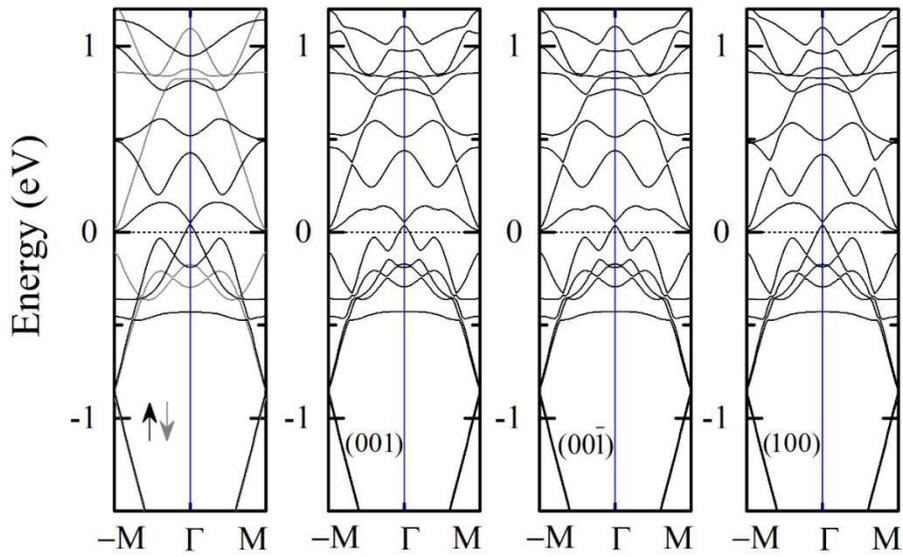

Figure S3. (a) The band structure in the first Brillouin zone(−M~Γ~M) of non-perturbed and perturbed with the (001), (100) and (00-1) magnetization for Gd-4ZGNR at $h_2$ site, respectively. The Fermi energy is set as zero.

## (e) Rashba parameter

The Rashba Hamiltonian is $H = \alpha\sigma(k_x \times e_z)$, here, $\alpha$ is the Rashba parameter which represents strength of Rashba effect, $\sigma$ is the Pauli matrix, $k_x$ the in-plane momentum, $e_z$ is the surface normal vector. Due to spin-orbital coupling, the spin splitting of band structure could be described by:

$$E^{\pm}(k) = \frac{\hbar^2}{2m^*}\vec{k}_x^2 \pm \alpha\vec{k}_x \quad (5)$$

where $m^*$ represents the electron effective mass.

The Eq. (5) can be written as

$$E^{\pm}(k) = \frac{\hbar^2}{2m^*}\left(k_x \pm \frac{m^*\alpha}{\hbar^2}\right)^2 - \frac{m^*\alpha^2}{2\hbar^2} \quad (6)$$

Thus, the momentum splitting of the Rashba bands from the crossing point is defined as $\Delta k_R = \frac{m^*\alpha}{\hbar^2}$. Finally, the Rashba parameter could be deduced as $\alpha = \hbar^2 \Delta k_R / m^*$.

## (f) The possible adsorption sites of Gd adsorbed *n*-ZGNRs

The possible adsorption hollow sites of Gd adsorbed *n*-ZGNR (*n*=5, 6, 7, 8, 9, 10, 12) are provide in Fig. S4. In addition, the energy differences among the antiferromagnetic (AFM), ferromagnetic (FM), nonmagnetic (NM) states of *n*-ZGNRs (*n*=5, 6, 7, 8, 9, 10, 12) are given as $E_{AFM-FM}$ and $E_{AFM-NM}$ in Table S1. Our calculated values are in good agreement with previous results reported by Krychowski et. al[1].

TABLE S1. Energy differences among the antiferromagnetic (AFM), ferromagnetic (FM), nonmagnetic (NM) states of *n*-ZGNRs (*n*=5, 6, 7, 8, 9, 10, 12). The results reported by Krychowski et. al[1] are also provided.

| *n* | $E_{AFM-FM}$ (meV) | | $E_{AFM-NM}$ (meV) | |
|---|---|---|---|---|
| | This work | Others[1] | This work | Others[1] |
| 5 | -10.85 | -10.67 | -64.32 | -65.10 |
| 6 | -11.68 | -11.39 | -68.56 | -69.41 |
| 7 | -10.33 | -8.83 | -77.62 | -71.64 |
| 8 | -7.735 | -5.88 | -72.83 | -73.73 |
| 9 | -6.78 | -4.23 | -78.59 | -75.75 |
| 10 | -6.42 | -3.34 | -75.837 | -77.47 |
| 12 | -2.59 | | -78.42 | |

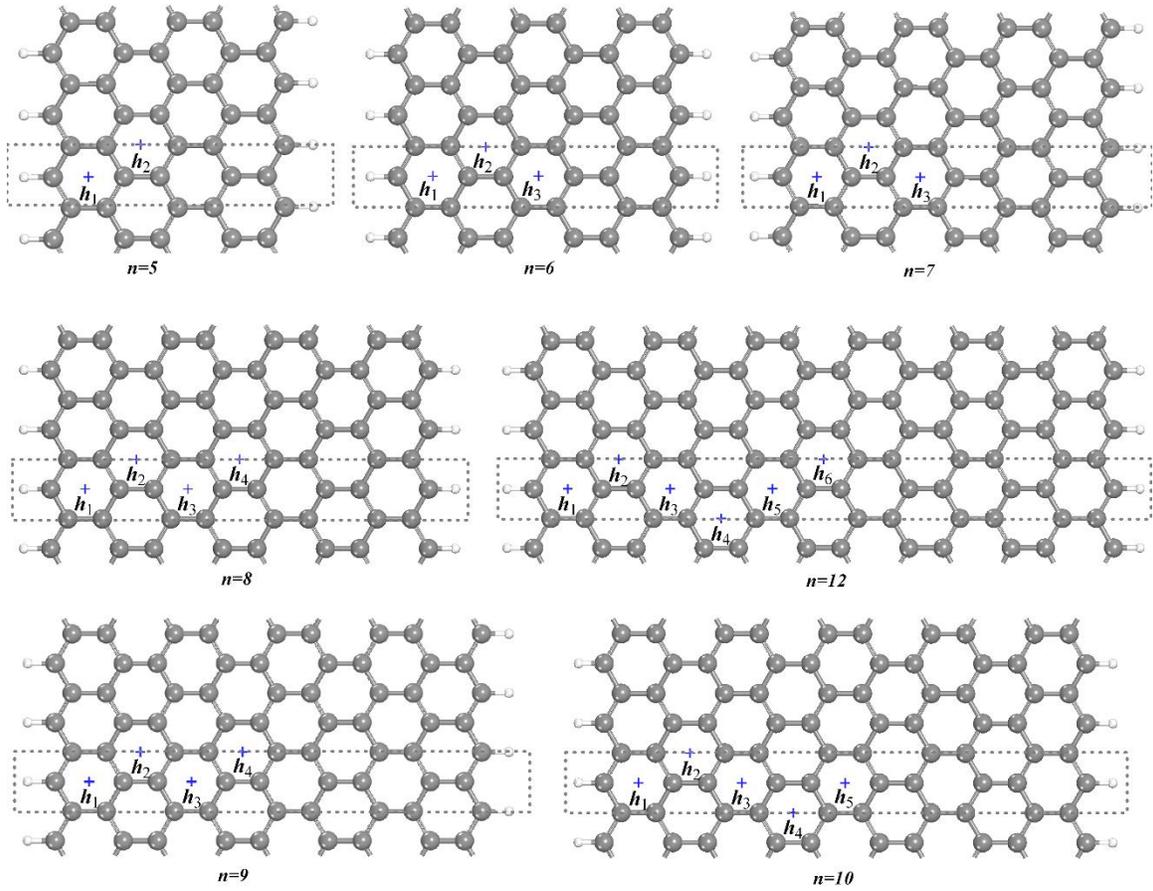

Figure S4. (a) Structures of *n*-ZGNRs with the width *n*= 5, 6, 7, 8, 9, 10, 12. The inequivalence possible hollow sites are labeled for Gd-*n*ZGNRs (*n*= 5, 6, 7, 8, 9, 10, 12) respectively. Such as, for the 12-ZGNR, there are six possible adsorption hollow sites ($h_1$, $h_2$, $h_3$, $h_4$, $h_5$, $h_6$).

# References

[1] D. Krychowski, J. Kaczkowski, and S. Lipinski, *Phys. Rev. B* **89**, 35424 (2014).